# Ethics-Based Auditing to Develop Trustworthy AI

Jakob Mökander[1] · Luciano Floridi[1,2]

**Abstract**
A series of recent developments points towards auditing as a promising mechanism to bridge the gap between principles and practice in AI ethics. Building on ongoing discussions concerning ethics-based auditing, we offer three contributions. First, we argue that ethics-based auditing can improve the quality of decision making, increase user satisfaction, unlock growth potential, enable law-making, and relieve human suffering. Second, we highlight current best practices to support the design and implementation of ethics-based auditing: To be feasible and effective, ethics-based auditing should take the form of a continuous and constructive process, approach ethical alignment from a system perspective, and be aligned with public policies and incentives for ethically desirable behaviour. Third, we identify and discuss the constraints associated with ethics-based auditing. Only by understanding and accounting for these constraints can ethics-based auditing facilitate ethical alignment of AI, while enabling society to reap the full economic and social benefits of automation.

**Keywords** Artificial intelligence · Auditing · Best practice · Ethics · Governance · Process

## 1 Towards trustworthy AI

The capacity to address the ethical challenges posed by artificial intelligence (AI) is quickly becoming a prerequisite for good governance (AI HLEG 2019). Unfortunately, the safeguards available to oversee human decision-making often fail

Jakob Mökander and Luciano Floridi have contributed equally to the article.

✉ Jakob Mökander
  jakob.mokander@oii.ox.ac.uk

  Luciano Floridi
  luciano.floridi@oii.ox.ac.uk

[1] Oxford Internet Institute, University of Oxford, 1 St Giles', Oxford OX1 3JS, UK

[2] Alan Turing Institute, British Library, 96 Euston Rd, London NW1 2DB, UK

          

when applied to AI. New mechanisms are thus needed to ensure ethical alignment of the AI systems that increasingly permeate society.

Leading institutions across the political, commercial, and academic strata of society have responded to the urgency of this task by creating ethics guidelines for trustworthy AI (Floridi and Cowls 2019). However, the adaptation of such guidelines remains voluntary. Moreover, the industry lacks useful tools and incentives to translate high-level ethics principles to verifiable and actionable criteria for designing and deploying AI (Raji et al. 2020).

A series of recent developments points towards ethics-based auditing as a promising mechanism to bridge the gap between principles and practice in AI ethics. A landmark article, published in April 2020 by leading researchers from, among others, Google, Intel, Oxford, Cambridge and Stanford, suggests that third-party auditors can be tasked with assessing whether safety, security, privacy, and fairness-related claims made by AI developers are accurate (Brundage et al. 2020). In parallel, professional services firms like PwC and Deloitte are developing frameworks to help clients design and deploy trustworthy AI (PwC 2019; Deloitte 2020).

We encourage this development. Nevertheless, it is important to remain realistic about what ethics-based auditing of AI can and cannot be reasonably expected to achieve.

## 2 Ethics-based Auditing of AI—What it is and How it Works

Ethics-based auditing is a governance mechanism that can be used by organisations that design and deploy AI systems to control or influence the behaviour of AI systems. Operationally, ethics-based auditing is characterised by a structured process by which an entity's behaviour is assessed for consistency with relevant principles or norms.

While AI should also be lawful and technically robust, our focus here is on the ethical aspects, i.e., what ought and ought not to be done over and above the existing regulation. Rather than attempting to codify ethics, ethics-based auditing helps identify, visualise, and communicate whichever normative values are embedded in a system.

Although standards have yet to emerge, a range of different approaches to ethics-based auditing of AI already exists: Functionality audits focus on the rationale behind the decision, code audits entail reviewing the source code, and impact audits investigate the effects of an algorithm's outputs.

Whether the auditor is a government body, a third-party contractor, or a specially designated function within larger organisations, the point is to ensure that the auditing runs independently of the day-to-day management of the auditee.

Ethics-based auditing contributes to good governance. Just as businesses require physical infrastructures to succeed, interactions between agents require an ethical infrastructure to flourish (Floridi 2014). By promoting procedural regularity and strengthening institutional trust, ethics-based auditing can help:

1. Provide decision-making support by visualising and monitoring outcomes
2. Inform individuals why a decision was reached and how to contest it
3. Allow for a sector-specific approach to AI governance
4. Relieve human suffering by anticipating and mitigating harms
5. Allocate accountability by tapping into existing governance structures
6. Balance conflicts of interest, e.g. by containing access to sensitive information to an authorised third-party.

## 3 Getting it Right

Ethics-based auditing of AI need not be difficult to implement. To be feasible and effective, however, specific requirements must be met. As a gold standard, we propose that the process of ethics-based auditing should be continuous, holistic, dialectic, strategic and design-driven.

First, ethics-based auditing is a process, not a destination. This implies that audits need to continuously monitor and evaluate system output and document performance characteristics.

Second, AI is not an isolated technology, but part of larger socio-technical systems. Thus, a holistic approach to ethics-based auditing takes knowledge of available alternatives into account when evaluating AI-based systems.

Third, ethics does not provide an answer sheet but a playbook. Ethics-based auditing should thus be viewed as a dialectical process wherein the auditor exists to ensure that the right questions have been asked.

Fourth, doing the right thing should be made easy. In practice, strategic value alignment means that ethics-based auditing frameworks must harmonise with organisational policies and individual incentives.

Finally, trustworthy AI is about design. Hence, interpretability and robustness should be built into systems from the start. Ethics-based auditing support this aim by providing active feedback to the continuous (re-)design process.

## 4 A Roadmap for Future Research

Ethics-based auditing of AI is not a panacea. In fact, as a governance mechanism, it is subject to a range of conceptual, technical, economic, social, organisational, and institutional constraints. These are listed in Table 1.

Conceptual constraints are logical limitations which cannot be easily resolved but need to be continuously managed. How to prioritise amongst incompatible definitions of concepts like fairness and justice, for example, remains a fundamentally political question. Hence, one function of ethics-based auditing would be to arrive at resolutions that, even when imperfect, are at least publicly defensible.

Technical constraints are tied to the autonomous, complex, and scalable nature of AI. For example, meaningful quality assurance of AI-based systems is not always possible within test environments, due to their ability to update their internal decision-making logic over time. However, since these constraints are context-dependent, they are likely to be relaxed or transformed by future research.

**Table 1** Constraints on auditing as a mechanism to ensure trustworthy AI

| Type | Constraints |
| --- | --- |
| Conceptual | There is a lack of consensus around high-level ethics principles |
|  | Normative values conflict and require tradeoffs |
|  | It is difficult to quantify externalities of complex AI systems |
|  | Information is infallibly lost through reductionist explanations |
| Technical | AI systems may appear opaque and can be hard to interpret |
|  | Data integrity and privacy are exposed to risks during audits |
|  | Linear compliance mechanisms are incompatible with agile software development |
|  | Tests may not be indicative of AI systems behaviour in real-world environments |
| Economic & social | Audits may disproportionately disadvantage or burden specific sectors or groups |
|  | Ensuring ethical alignment must be balanced with incentives for innovation |
|  | Ethics-based auditing is vulnerable to adversarial behaviour |
|  | The transformative effects of AI pose challenges for how to trigger audits |
|  | Ethics-based auditing may reflect and reinforce existing power structures |
| Organisational & institutional | There is a lack of institutional clarity about who audits whom |
|  | Auditors may lack the access or information required to evaluate AI systems |
|  | The global nature of AI systems challenge national jurisdictions |

Economic and social constraints depend on the incentives of different actors. Because ethics-based auditing imposes costs, financial and otherwise, care must be taken to not unduly burden particular sectors or groups in society. At the same time, effective governance cannot afford to be naïve. Even in cases where audits reveal flaws in AI-based systems, asymmetries of power may prevent corrective steps from being taken.

Organisational constraints concern the design of operational auditing frameworks. Ethics-based auditing is only as good as the institutions backing it. Currently, a clear institutional structure is lacking. Moreover, the effectiveness of ethics-based auditing remains constrained by a tension between national jurisdictions, on the one hand, and the global nature of technology, on the other.

## 5 Outlook

Policymakers are encouraged to consider ethics-based auditing as an integral component of holistic approaches to managing the ethical risks posed by AI. This does not mean that traditional compliance mechanisms are redundant. Instead, ethics-based auditing of AI holds the potential to complement and enhance other tools and methods like human oversight, certification, and regulation.

For ethics-based auditing to be effective, the shift of power from juridical courts to private actors must be resisted. However, too strict laws risk stifling innovation and undermining the legitimacy of law enforcement. The solution here is that governments retain supreme sanctioning power by authorising independent agencies who, in turn, conduct ethics-based audits of AI systems.

Much remains to be done. Future research should focus on the 16 constraints listed in Table 1. Only if these constraints are understood and accounted for will ethics-based auditing help address the ethical risks posed by AI.

In conclusion, ethics-based auditing of AI affords good governance by strengthening the ethical infrastructure of mature information societies. However, ethics-based auditing will not and should not replace the need for continuous ethical reflection among individual moral agents.

**Funding** Jakob Mökander was supported by the Wallenberg Foundation.